# Tunneling spectroscopy and magnetization measurements of the superconductor properties of MgB$_2$

A. Sharoni, I. Felner*, and O. Millo**

*Racah Institute of Physics, The Hebrew University, Jerusalem 91904, Israel*

**Abstract**

Cryogenic scanning tunneling microscopy and magnetization measurements were used to study the superconducting properties of MgB$_2$. The magnetization measurements show a sharp superconductor transition at T$_C$ = 39 K, in agreement with previous works. The tunneling spectra exhibit BCS-like gap structures, with gap parameters in the range of 5 to 7 meV, yielding a ratio of 2$\Delta$/$K_B T_C$ ~ 3-4. This suggests that MgB$_2$ is a conventional BCS *s*-wave superconductor, either in the weak-coupling or in the `intermediate-coupling` regime.

* e-mail: milode@vms.huji.ac.il,   **e-mail: israela@vms.huji.ac.il

The recent discovery (1) of superconductivity at T$_C$ ~ 39 K in the simple intermetallic compound MgB$_2$ is particularly surprising for many reasons. This T$_C$ is much higher than the highest T$_C$ values reported for any other intermetallic compound, such as the A15 compounds (Nb$_3$Ge, T$_C$ = 23.2 K) and the borocarbides (YPd$_2$B$_2$C, T$_C$=23 K), as well as for any non-oxide and non C$_{60}$-based material. Moreover, this transition temperature is above the limit predicted theoretically for conventional BCS superconductivity in the weak-coupling regime (2). Therefore, a question arises regarding the mechanism for superconductivity in this system. The Boron isotope effect measured by Bud`ko et al. (3) suggests that MgB$_2$ is a BCS phonon-mediated superconductor and also indicates that the high transition temperature is partly due to the low mass of boron that yields high phonon frequencies.

Further evidence for BCS superconductivity can be provided by tunneling spectroscopy, which yield direct information on the quasi-particle density of states (DOS) (4). Tunneling spectroscopy measurements on MgB$_2$ performed by Rubio-Bollinger et al. show a BCS quasi-particle DOS on small MgB$_2$ grains embedded in a gold matrix (5). However, the transition temperature was not reported for their samples and the observed gap parameter, $\Delta$ = 2 meV, correspond via the BCS relation 2$\Delta$/$K_B T_C$ = 3.53 (4) to a transition temperature of 13.2 K. Karapetrov et al. have presented tunneling spectroscopy data yielding superconductor gaps that conform with the weak-coupling BCS theory (6), although their tunneling spectra exhibited relatively large smearing. Schmidt et al. performed point-contact measurements that exhibited a BCS gap structure with $\Delta$ = 4.3 meV, smaller than the BCS prediction. This apparently smaller value was attributed to a chemically modified surface layer (7). On the other hand, NMR (8) and Raman spectroscopy (9) measurements indicated a bulk gap value which is significantly larger than the BCS prediction. Here, we present a combined magnetization and tunneling spectroscopy study of MgB$_2$ with T$_C$ = 39 K, as determined by our magnetization measurements. The tunneling spectra manifest BCS-like gap structures with gap parameters ranging between 5 to 7 meV, around the weak coupling BCS prediction for this T$_C$, $\Delta$ ~ 5.9 meV.

Stoichiometric ratio of Mg and B elements (99.9% pure) in lump form were placed in a Ta tube. The Ta tube was then sealed in an evacuated quartz ampoule and heated to 950 C in a box furnace for two hours. The powder X-ray diffraction pattern was indexed to the well known hexagonal $AlB_2$-type unit cell of $MgB_2$, which has been structurally characterized already in the mid 1950's (10). This structure can be viewed as an intercalated graphite structure with full occupation of interstitial sites centered in hexagonal prism made of B atoms. The lattice parameters obtained, a=3.110 and c=3.519 Å, are in excellent agreement with the data given in Refs. (11) and (12). The pattern contained a few extra peaks (with intensity of less than 5%) which are due to boron, possibly resulting from a small evaporation of Mg in the growth process.

The dc magnetic measurements on solid ceramic pieces in the range of 5-45 K were performed in a commercial (Quantum Design) superconducting quantum interference device magnetometer (SQUID). Figure 1 shows zero-field-cooled (ZFC) and field-cooled (FC) magnetization curves of $MgB_2$ measured at 2 Oe. From this figure we obtain $T_C$=39 K, and the relatively low values of the FC branch indicate strong flux pinning, thus suggesting a very good possibility of high current superconducting applications for this material.

Figure 2 shows the isothermal magnetization M(H) curves of $MgB_2$ measured at various temperatures between 5 and 35 K. The curves are symmetric with respect to increasing and decreasing of the applied fields, typical to type II superconductors (for the sake of brevity only the positive parts are shown). From these curves we can estimate the lower critical field $H_{C1}(0)$. First, $H_{C1}(T)$ were determined as the magnetic fields at which a deviation from linearity of the virgin magnetization curves were observed. The obtained $H_{C1}(T)$ values are plotted in Fig 3, and an extrapolation of the data to zero field yields the value for $H_{C1}(0)$, ~ 450 Oe. The sample was nearly spherical in shape so no demagnetizing factor was necessary. From the hysteresis of the magnetization curves, $\Delta M$(emu/cc), we have estimated the critical current density $J_C$ as a function of the external applied fields at various temperatures, assuming the critical state model (the Bean model). The values obtained are consistent with data reported in other papers (11).

For the tunneling measurements the samples were cleaned with distilled water in an ultrasonic bath and then dried using highly pure nitrogen just before mounting in our homemade cryogenic STM. Some of the samples were polished first by mettalographic grinding paper and ending with 0.25 μm diamond lapping-compound. This polishing procedure had no detectable influence on our spectroscopic data. The STM was immersed in liquid He right after evacuating the sample-space, and the sample and the scan-head were cooled down to 4.2 K via He exchange gas. The tunneling spectra, namely, the I-V and the tunneling conductance dI/dV vs. V curves (the latter is proportional to the local DOS), were acquired while momentarily disconnecting the feedback circuit. We note that the tunneling conductance curves measured directly using conventional lock-in technique were nearly identical to those obtained by numerical differentiation of the I-V curves. At each position we have checked the possibility that the gaps in the spectra originated from the Coulomb blockade and not from superconductivity. This was done by acquiring spectra at different settings of the tunneling current and sample-tip bias. The STM setting affects the Coulomb gap, which may even vanish for particular settings (13), but not the gap arising from superconductivity (14).

In Fig. 4 we present four tunneling I-V and dI/dV vs V characteristics, representative of many spectra acquired on different samples and at different locations, showing minimal (a), mid-range (b,c) and maximal (d) gaps observed in our data. These spectra exhibit a small asymmetry, but can still be fit relatively well to the Dynes function for tunneling into a BCS (*s*-wave) superconductor (15), with gap parameters $\Delta$ = 5, 5.2, 5.5 and 7 meV for spectra (a)-(d), respectively. The lifetime broadening parameters used in these fits were relatively small, about $0.1\Delta$. With the measured $T_C$ = 39 K we obtain that the ratio $2\Delta/k_B T_C$ varies between 3.0 to 4.2, around the theoretical value (3.53) for a weak-coupling BCS superconductor. The gap values were not evenly distributed in this range, and most of the surface (~ 80%) showed gaps between 5 to 6 meV ($2\Delta/k_B T_C$ between 3 to 3.6). We note that while on most of the surface a superconductor gap structure was clearly observed (such as shown in Fig. 4), we found some regions where the gap was highly suppressed or even vanished. This can be attributed to local defects or to the existence of a minor non-superconducting impurity phase.

Our tunneling data suggest that $MgB_2$ is a BCS-type *s*-wave superconductor. We found no clear evidence for *d*-wave superconductivity in the tunneling spectra, such as zero bias conductance peaks (16, 17). Although the measured spectra exhibited, in many cases, some asymmetry, this effect was much smaller than what is typically observed in the high $T_C$ cuprates (16, 18). Such a small asymmetry was predicted to exist also for *s*-wave superconductors having specific band structures (19). However, in spite of the discussion above, we think that *d*-wave superconductivity or any other order parameter symmetry cannot yet be ruled out. Measurements on clean single crystals are needed for an unambiguous determination of the order parameter symmetry.

The spatial variation in the measured superconducting gaps described above does not allow us to determine whether or not $MgB_2$ is in the weak-coupling regime, for which $2\Delta/k_B T_C \approx 3.5$. Assuming that the maximal gaps, for which $2\Delta/k_B T_C$ = 4.2 (observed on a small fraction of the sample surface), were measured at locations where the surface material is of the highest purity, then $MgB_2$ may possibly be in the `intermediate coupling` regime. This is consistent with the NMR and Raman measurements mentioned above (8, 9). On the other hand, the local enhancement of $\Delta$ may result from an effect similar to that causing the enhancement of $T_C$ in granular weak-coupling superconductors (20). This latter scenario is consistent with a picture of $MgB_2$ being in the weak-coupling regime. We note that the gaps measured in our experiment are larger than typically measured in other tunneling experiments (5, 7), although Karapetrov et al (6) reported a value of 5.2 meV, which is within the range observed by us. A distribution of smaller gaps (4-5.2 meV) as well as spectra exhibiting in-gap states (but no zero-bias conductance peaks) were recently observed by us in measurements performed on a $MgB_2$/Al composite (21). This observation may be due to the proximity effect or surface chemical modification, as noted by Schmidt et al. (7). The variance in the gap values and gap structures observed in measurements performed on different $MgB_2$ samples prepared in different ways (59, 21) emphasizes the need for performing measurements on pure single-crystal materials. There, one can look for possible directional dependence of the tunneling spectra, which may shed light on the origin of these variations and, more important, on the coupling mechanism and symmetry of the order parameter in $MgB_2$.

In summary, we present magnetization measurements confirming that $MgB_2$ is a type II superconductor with $T_C$ ~ 39 K, and $H_{C1}(0)$ ~ 450 Oe. These data also indicate the existence of strong flux pinning, making this material a good candidate for high-current applications. Tunneling spectra measured on $MgB_2$ reveal a BCS-like quasi-particle DOS, with gap parameters that vary spatially between 5 to 7 meV, thus $2\Delta/k_BT_C$ varies between 3.0 to 4.2. No clear evidence for *d*-wave symmetry of the order parameter was found in the tunneling spectra. Our tunneling data thus suggest that MgB$_2$ is a BCS *s*-wave superconductor that may be either in the weak-coupling or in the `intermediate-coupling` regime.

This research was supported by the BSF (1998), by the Klachky Foundation, and by the Israel Academy of Sciences.

**Figures and Captions:**

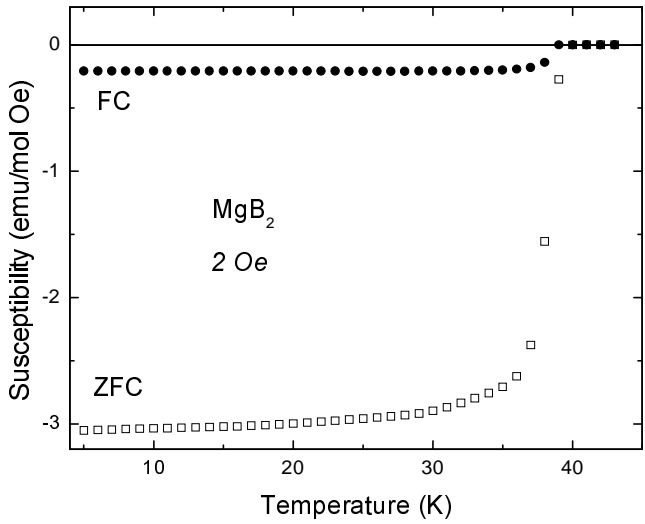

**Fig. 1** ZFC and FC magnetization curves for $MgB_2$ measured at 2 Oe.

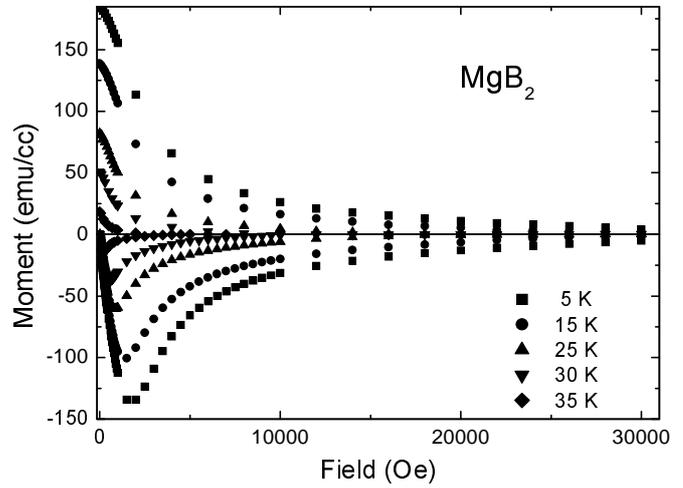

**Fig. 2** Isothermal magnetization curves at various temperatures for $MgB_2$.

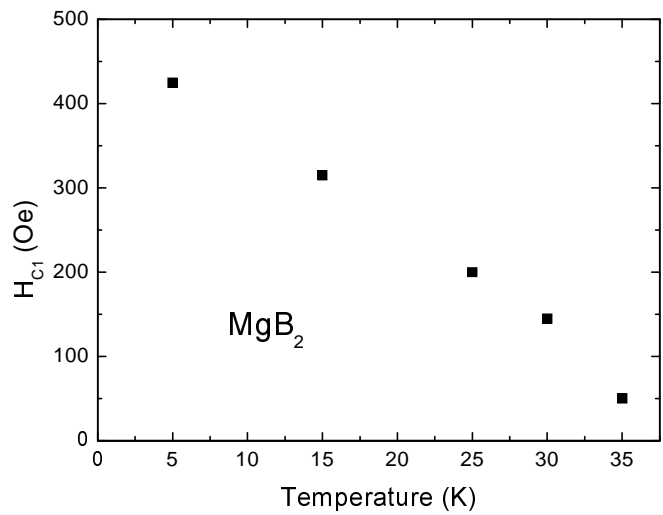

**Fig. 3** Temperature dependence of the lower critical field $H_{C1}$, determined as the field at which a deviation from linearity of the virgin magnetization curve M(H) shown in Fig. 2 are observed.

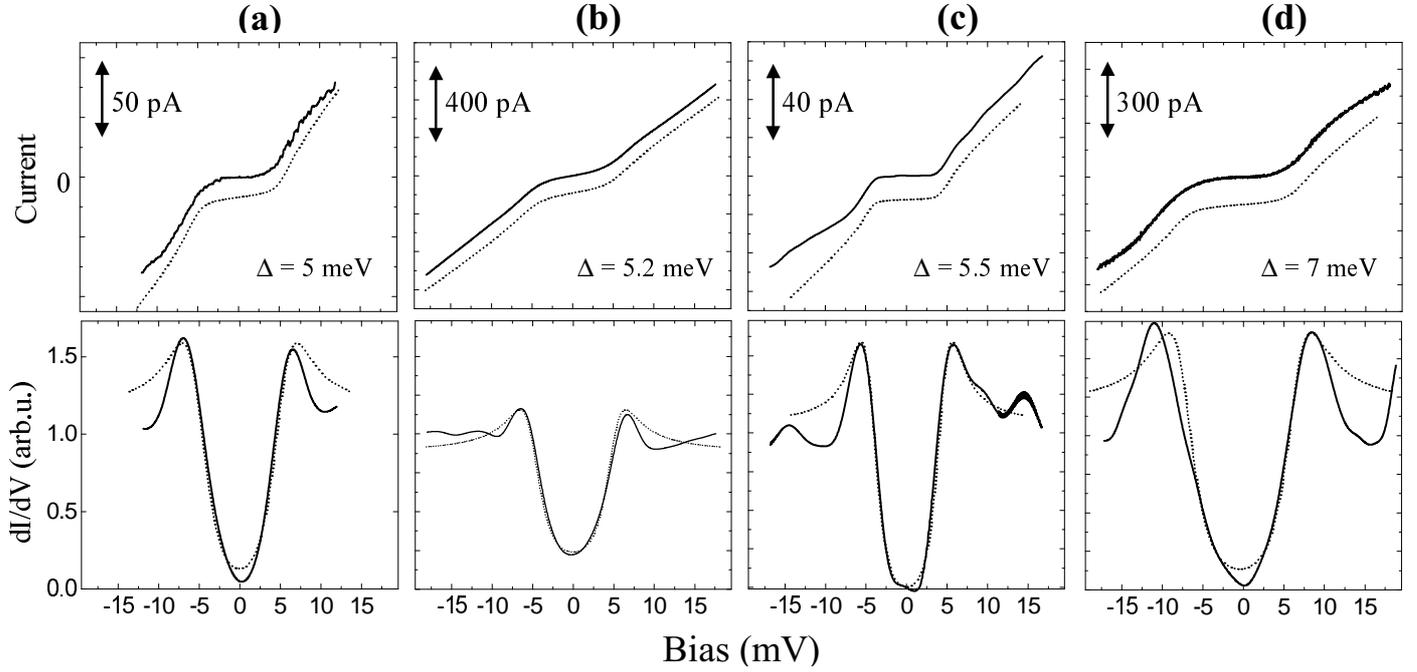

**Fig. 4**. I-V charateristics (upper frames) and dI/dV-V tunneling conductance spectra (lower frames) taken at 4.2 K at different locations on the surface of $MgB_2$. The solid curves represent measured data, while the dotted curves are the fits, shifted for clarity in the I-V curves. The tunneling conductance spectra are normalized to the normal junction conductance at a bias larger than the superconductor gap, and the current scales for the I-V curves are denoted in the figures. (a) shows a spectrum having the smallest measured gap, $\Delta = 5$ meV, (b) and (c) show midrange gap structures, $\Delta = 5.2$ and 5.5 meV, respectively, and (d) has the largest observed gap, $\Delta = 7$ meV. All fits required a Dynes broadening parameter of ~ $0.1\Delta$.